\documentclass[a4paper,fleqn]{cas-dc}

\usepackage[utf8]{inputenc}
\usepackage[square,sort&compress,numbers]{natbib}
\usepackage{graphicx}
\usepackage{epsfig}

\usepackage{caption}
\usepackage{subcaption}
\usepackage{siunitx}
\usepackage[capitalise]{cleveref}
\usepackage[version=4]{mhchem}
\usepackage{physics}

\DeclareSIUnit\angstrom{\text{\AA}}
\newcommand{\hystTemp}{\SI[separate-uncertainty=true]{274 \pm 11}{\kelvin}}
\newcommand{\hystCenter}{\SI{274}{\kelvin}}
\newcommand{\hystWidth}{\SI{22}{\kelvin}}
\newcommand{\muSR}{\(\mu^{+}\)SR}

\begin{document}
\let\WriteBookmarks\relax
\def\floatpagepagefraction{1}
\def\textpagefraction{.001}

\title{Muon-spin relaxation investigation of magnetic bistability in a crystalline
  organic radical compound }
\shorttitle{Muon-spin relaxation investigation of magnetic bistability}
\author[1]{A. Hern\'{a}ndez-Mel\'{i}an}
\cormark[1]
\ead{alberto.hernandez-melian@durham.ac.uk}
\cortext[1]{Corresponding author}
\author[1]{B.~M. Huddart}
\author[2]{F.~L. Pratt}
\author[3]{S.~J. Blundell}
\author[4]{M. Mills}
\author[4]{H.~K.~S. Young}
\author[4]{K.~E. Preuss}
\author[1]{T. Lancaster}
\shortauthors{A. Hern\'{a}ndez-Mel\'{i}an}

\affiliation[1]{
  organization={Department of Physics, Centre for Materials Physics, Durham University},
  addressline={South Road},
  city={Durham},
  postcode={DH1 3LE},
  country={United Kingdom}
}

\affiliation[2]{
  organization={ISIS Neutron and Muon Source},
  addressline={STFC-RAL},
  city={Chilton, Didcot},
  postcode={OX11 OQX},
  country={United Kingdom}
}

\affiliation[3]{
  organization={Oxford University},
  addressline={Clarendon Laboratory, Parks Road},
  city={Oxford},
  postcode={OX1 3PU},
  country={United Kingdom}
}

\affiliation[4]{
  organization={Department of Chemistry, University of Guelph},
  addressline={50 Stone Road East},
  city={Guelph},
  postcode={N1G 2W1},
  state={Ontario},
  country={Canada}
}

\begin{abstract}
  We present the results of a muon-spin relaxation (\muSR)
  investigation of the crystalline organic radical compound
  4-(2-benzimidazolyl)-1,2,3,5-dithiadiazolyl (HbimDTDA), in which we
  demonstrate the hysteretic magnetic switching of the system that
  takes place at \(T = \hystTemp\) caused by a structural phase
  transition. Muon-site analysis using electronic structure
  calculations suggests a range of candidate muon stopping sites. The
  sites are numerous and similar in energy but, significantly, differ
  between the two structural phases of the material. Despite the
  difference in the sites, the muon remains a faithful probe of the
  transition, revealing a dynamically-fluctuating magnetically
  disordered state in the low-temperature structural phase. In
  contrast, in the high temperature phase the relaxation is caused by
  static nuclear moments, with rapid electronic dynamics being
  motionally narrowed from the muon spectra.
\end{abstract}
\maketitle

\section{Introduction}

\begin{figure}
  \centering
  \begin{subfigure}{0.60\columnwidth}
    \includegraphics[width=\linewidth]{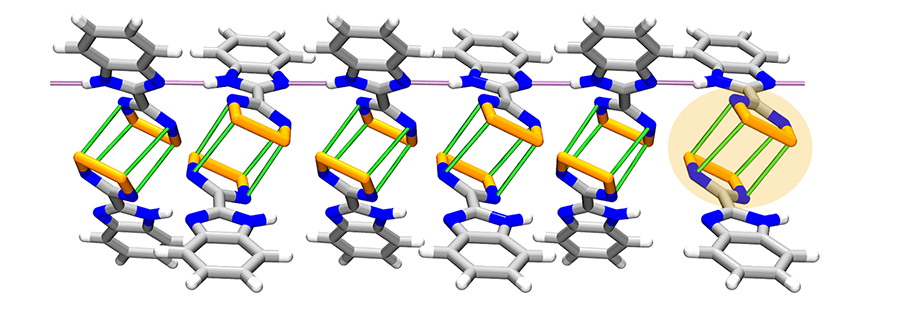}
    \subcaption{\SI{100}{\kelvin}}
    \label{fig/molecule:chain_low}
    \includegraphics[width=\linewidth]{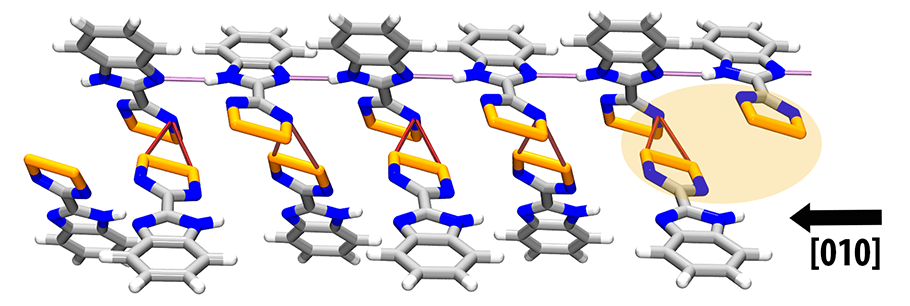}
    \subcaption{\SI{340}{\kelvin}}
    \label{fig/molecule:chain_high}
  \end{subfigure}
  \begin{subfigure}{0.30\columnwidth}
    \includegraphics[width=0.75\linewidth]{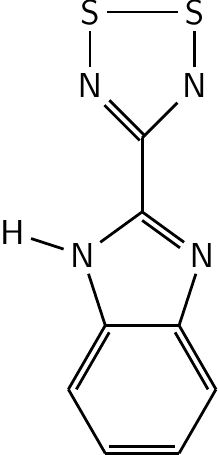}
    \subcaption{HbimDTDA}
    \label{fig/molecule:chemfig}
  \end{subfigure}
  \caption{
    (\subref*{fig/molecule:chain_low}) Low-temperature structure of HbimDTDA, with pancake bonds between the linear arrays of molecules arranged along \([010]\).
    (\subref*{fig/molecule:chain_high}) High-temperature structure following a shift along \([010]\) that breaks the pancake bonds.
    (\subref*{fig/molecule:chemfig}) Chemical structure diagram of single HbimDTDA molecule.}
  \label{fig/molecule}
\end{figure}

Understanding the link between hysteresis and structure is an
important theme in materials design, since systems that exhibit
hysteretic effects intrinsically possess memory and are therefore of
potential technological interest. Recently, a crystalline organic
radical compound 4-(2-benzimidazolyl)-1,2,3,5-dithiadiazolyl
(HbimDTDA) was reported \cite{mills2018} that exhibits bistability in
its magnetic and structural properties near room temperature. In the
solid state, the neutral radical HbimDTDA crystallizes in an
orthorhombic \(Pbca\) space group (see \cref{fig/molecule}). The
magnetic switching effect follows from a subtle
single-crystal-to-single-crystal structural phase transition centred
at \(T \approx \SI{270}{\kelvin}\) that occurs without symmetry
breaking, but involves a significant reorganisation of supramolecular
contacts. Structural analysis at \(T = \SI{100}{\kelvin}\) shows that
the low-temperature structure of the material involves one-dimensional
linear arrays of HbimDTDA molecules, with each molecule forming part
of a pancake-bonded pair with a partner molecule on a neighbouring
array [\cref{fig/molecule:chain_low}]. The geometry of the pancake
bonds, determined by overlap of the four lobes of each molecule's
singly-occupied molecular orbital, orients the molecules to create a
dense 3D network of supramolecular contacts. In contrast, the
high-temperature structure of the system determined at
\(T = \SI{340}{\kelvin}\) does not feature the pancake bonds, which
are broken and replaced with new electrostatic contacts
[\cref{fig/molecule:chain_high}]. These two structural phases are
related by a translation in the \([010]\) direction, such that the
one-dimensional supramolecular structures (defined by hydrogen bonding
between neighbouring molecules) shift with respect to one another.
Analysis of the temperature dependence of the structural phase
transition confirms a first-order transition between two unique
phases, occurring around \(T\approx \SI{270}{\kelvin}\), with
significant thermal hysteresis \cite{mills2018}.

Each radical unit carries a \(S = 1/2\) spin and the magnetism of the
system is closely linked to the structural transition. Magnetic
susceptibility data were reported to indicate diamagnetic
low-temperature behaviour, which was explained by the electronic
overlaps promoted by the pancake bonds between \(\pi\)-radicals. At
high temperature the susceptibility increases dramatically, consistent
with the non-pancake bonded phase being paramagnetic and comprising an
unpaired \(S = 1/2\) spin per molecule, with some degree of
antiferromagnetic coupling between them. Particularly notable is the
hysteretic magnetic transition between the states \cite{mills2018}.

Implanted muons are widely used as local probes of magnetism
\cite{blundell2022}, with their extreme sensitivity motivating their
use to determine the magnetic order and dynamics in low-moment,
molecule-based magnets. Muons have been used rather less to look at
magnetic bistability, though they have proved an effective probe of
molecular spin-crossover (SCO) materials formed from bistable
molecules that are able to switch from low to high spin states via a
cooperative phase transformation \cite{blundell2003}. In this paper we
report the use of muon-spin relaxation (\muSR) techniques to examine
the cooperative magnetic switching in HbimDTDA from a local
perspective. We show that muons are sensitive to the bistability of
the magnetic state and use this to elucidate the nature of the low-
and high-temperature regimes, and provide a determination of the
characteristic field fluctuation rate in the low-temperature regime.
We also determine the muon sites using first-principles electronic
structure methods to demonstrate how the muon is sensitive to the
magnetic environment in this chemically-complex material.

\section{Experimental}

In a \muSR\ experiment positive muons are implanted in the sample,
usually settling at interstitial sites, but quickly decay with a mean
lifetime of \(\SI{2.2}{\micro\second}\). The muon spin polarisation,
whose time-dependence is determined by the local field distribution,
can be measured by studying the statistics of the positron emission in
the decay, since the positron is emitted preferentially along the
direction of the muon spin. The detectors around the sample are
classified into forward (F) and backward (B) banks with respect to the
initial muon polarisation, so that the quantity of interest,
proportional to the muon spin polarisation, is the positron asymmetry
function
\begin{equation}
  A(t) = \frac{N_{\mathrm{F}}(t) - \alpha N_{\mathrm{B}}(t)}{N_{\mathrm{F}}(t) + \alpha N_{\mathrm{B}}(t)},
\end{equation}
where \(N_{\mathrm{F}}(t)\) and \(N_{\mathrm{B}}(t)\) are sums of the
counts in all the forward and backward detectors respectively, while
\(\alpha\) is a calibration constant which accounts for the different
efficiencies and geometries of the detectors and can be determined
from experimental data.

To investigate the hysteresis effect and the magnetism of the two
phases of the compound, \muSR\ measurements were performed using the
HiFi spectrometer at the STFC-ISIS Facility (Rutherford Appleton
Laboratory, UK). We employed the longitudinal field (LF) geometry
where an external magnetic field is applied along the initial
muon-spin direction. Initially, a series of measurement were made in
zero applied magnetic field, sweeping temperatures such that each
measurement was made at a fixed temperature for \(\SI{35}{\minute}\),
with temperature changes taking \(\SI{7}{\minute}\). Measurements were
also made as a function of applied field at fixed temperature, for
\(\SI{40}{\minute}\) per point. We also performed weak transverse
field (wTF) measurements, where a small magnetic field
(\SI{2}{\milli\tesla}) is applied perpendicular to the initial
muon-spin direction. Each measurement took \(\SI{24}{\minute}\), with
\(\SI{7}{\minute}\) for temperature adjustment. A polycrystalline
sample of HbimDTDA was prepared as described previously
\cite{mills2018}. For the measurement it was wrapped in Ag foil,
sealed in an airtight Cu holder and then loaded into a \(\ce{^{4}He}\)
cryostat.

\section{Results}

\subsection{Zero-field measurements}

\begin{figure}
  \centering
  \includegraphics[width=0.7\columnwidth]{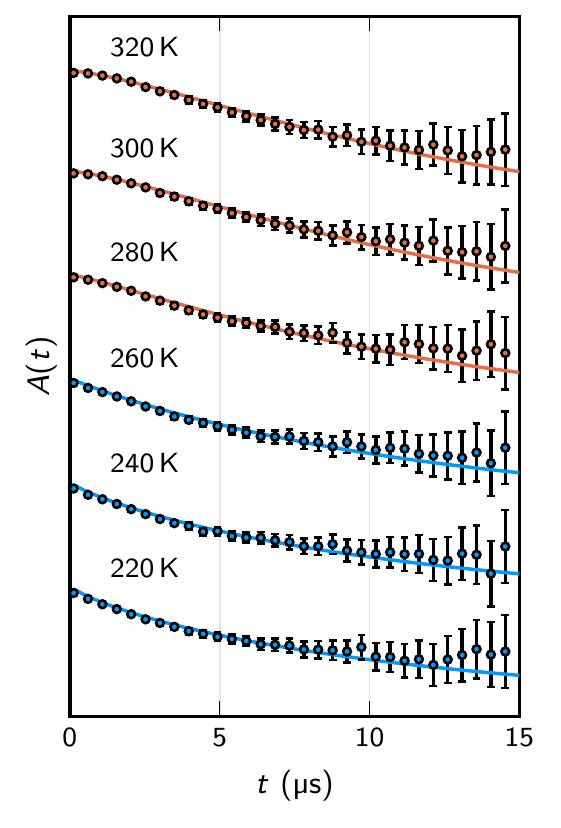}
  \caption{ZF asymmetry spectra measured on HbimDTDA at increasing
    temperatures across the transition, offset for clarity and
    therefore in arbitrary units.}
  \label{fig/zf_asy}
\end{figure}

The sample was first cooled to \(T = \SI{220}{\kelvin}\) and a series
of measurements in zero-applied field (ZF) were made in intervals of
\(\SI{10}{\kelvin}\) up to \(\SI{350}{\kelvin}\). Measurements were
then repeated for decreasing temperature. Example spectra for
measurements taken on increasing temperature are shown in
\cref{fig/zf_asy}. The observed trend is that spectra resemble an
exponential relaxation at low temperatures and become more Gaussian in
character as the temperature increases. To track their evolution, the
spectra were fitted to a stretched exponential relaxation function
\begin{equation}
  \label{eq/zf_model}
  A(t) = A_{\mathrm{R}}^{\mathrm{ZF}}\exp[-\qty(\lambda^{\mathrm{ZF}} t)^{\beta}] + A_{\mathrm{B}}^{\mathrm{ZF}},
\end{equation}
where the final term \(A_{\mathrm{B}}^{\mathrm{ZF}}\) accounts for
muon spins that do not relax, including those from muons implanted in
the sample holder. To simplify the fitting procedure we fix the
parameters which vary the least in a free fit, in this case
\(A_{\mathrm{B}}^{\mathrm{ZF}} = \SI{12}{\percent}\) and
\(\lambda^{\mathrm{ZF}} = \SI{0.08}{\per\micro\second}\) by taking an
average. We also find that the relaxing asymmetry
\(A_{\mathrm{R}}^{\mathrm{ZF}}\) increases from
\(\SI{15.5}{\percent}\) to \(\SI{16.6}{\percent}\) between the low-
and high-temperature phases. The parameter \(\beta\) allows us to
interpolate between an (i) approximately exponential decay, which
results from a combination of dynamically fluctuating, disordered
magnetic moments in the fast fluctuation limit; and (ii) behaviour
approaching Gaussian decay, which approximates the initial relaxation
of the Kubo-Toyabe function, which results from static magnetic
moments sampled from a normal distribution \cite{blundell2022}.

\begin{figure}
  \centering
  \includegraphics[width=0.95\columnwidth]{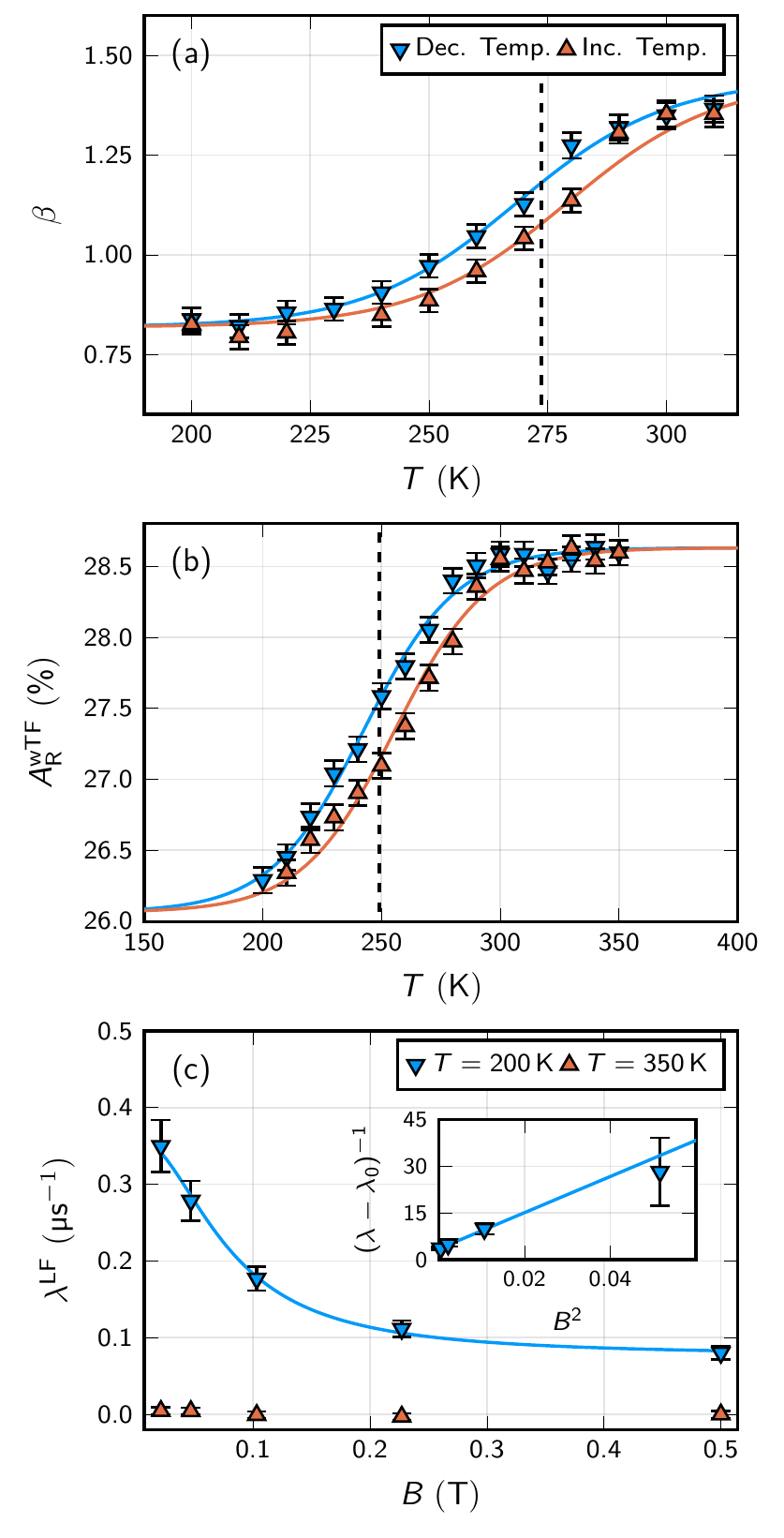}
  \caption{
    (a) The result of fitting a stretched exponential function [\cref{eq/zf_model}] to the ZF results, showing the temperature dependence of the line-shape parameter \(\beta\) for increasing (Inc.) and decreasing (Dec.) temperature.
    (b) The result of fitting an exponentially decaying cosine curve [\cref{eq/wtf_model}] to the wTF results, for which we show the relaxing asymmetry \(A_{\mathrm{R}}^{\mathrm{wTF}}\).
    (c) The result of fitting an exponential decay [\cref{eq/lf_model}] to the field-dependent LF data, giving the relaxation parameter \(\lambda^{\mathrm{LF}}\) fitted to the Redfield formula [\cref{eq/redfield_formula}].
  }
  \label{fig/mag_params}
\end{figure}

The results of the fitting procedure are shown in
\cref{fig/mag_params}, where the fitted stretching parameter is seen
to change as a function of temperature across a transition region
[\cref{fig/mag_params}(a)]. We can clearly see the hysteresis effect
with the decreasing-temperature measurements (down triangles)
consistently at higher values than the increasing-temperature ones (up
triangles) over a region centred on \(T = \hystTemp\). This value was
extracted from the fitted values of \(\beta\) by fitting both sets of
measurements to the phenomenological functional form
\begin{equation}
  A_{\mathrm{R}}^{\mathrm{wTF}}(T) = A_{\mathrm{H}} \tanh[k_{\mathrm{H}} (T - T_0)] + c_{\mathrm{H}},
\end{equation}
where \(A_{\mathrm{H}}\), \(k_{\mathrm{H}}\) and \(c_{\mathrm{H}}\)
are parameters which determine the shape and position of each curve
and are kept constant between them whilst \(T_0\) determines the
centre and is different between the increasing and decreasing
temperature curves. We can therefore determine an approximate value
for the width of transition by taking the difference between the
\(T_0\) values.

The difference between the two regimes can be explained by the
different distributions of fluctuating magnetic moments in each. In
the low temperature phase we have randomly-oriented electronic moments
(in a distribution of width
\(\Delta/\gamma_{\mu} = \sqrt{\langle B^{2}\rangle}\), where
\(\gamma_{\mu} = 2\pi \times \SI{135.5}{\mega\hertz\per\tesla}\) is
the muon gyromagnetic ratio) fluctuating at rate \(\nu\) in the fast
fluctuation limit \(\nu \gg \Delta\). As the temperature increases
through the transition, the density of moments increases due to the
structural transition. Crucially, these moments fluctuate at a much
faster rate in the higher temperature phase, with the result that the
muon spin, whose evolution is limited by the value of its gyromagnetic
ratio, cannot complete a rotation before the local field fluctuates
and changes value \cite{blundell2022}. The electronic moments are
therefore motionally narrowed from the spectra in the high-\(T\)
regime. This leaves only the random nuclear moments to account for a
large part of the relaxation. The nuclear spins are quasistatic and so
are described by a Kubo-Toyabe-like function (of which we only observe
the early-time, Gaussian part). The fact that the value of $\beta$
appears to plateau below $\beta \approx 1.5$ suggests that the
motional narrowing is not complete.

\subsection{Weak transverse-field measurements}

In order to confirm the existence of the hysteresis loop, the
temperature-dependent measurements were also repeated over the same
range but in a weak transverse magnetic field
(\(B = \SI{2}{\milli\tesla}\)). Since the external field is so low,
the only muon spins that will oscillate are those in sites where the
local field almost vanishes in zero field, and which are not rapidly
relaxed by dynamics. The size of the change we observe in amplitude
with temperature is small, suggesting that those muons contributing to
this effect constitute only a small fraction of the total ensemble,
which might be explained by the change in the nature of the muon sites
with structural phase, as discussed below. The results were fitted to
a decaying cosinusoidal curve
\begin{equation}
  \label{eq/wtf_model}
  A(t) = A_{\mathrm{R}}^{\mathrm{wTF}}\exp(-\lambda^{\mathrm{wTF}} t)\cos(\omega t + \phi) + A_{\mathrm{B}}^{\mathrm{wTF}},
\end{equation}
with the resulting relaxation asymmetry shown in
\cref{fig/mag_params}(b). We again see a consistent separation between
the measurements made in increasing and decreasing temperature over
the transition region, but compared to those above, the fitted
parameters have a much lower uncertainty (in part because fitting a
periodic cosine wave has less margin of error than an exponential),
and so the hysteresis loop is clearer. Repeating the fitting procedure
used in the previous section, we find that the loop for these
measurements is centred on the slightly lower temperature of
\(T = \SI[separate-uncertainty=true]{249 \pm 13}{\kelvin}\). The
discrepancy between this and the transition derived from the change in
the $\beta$ parameter suggests the two measurements are sensitive to
different aspects of the muon's interaction with the system: $\beta$
reflects the distribution of local magnetic fields;
$A_{\mathrm{R}}^{\mathrm{wTF}}$ reflects the availability of muon
sites in the two regimes, as described below. We note also that for
both sets of measurements the transition appears to be continuous,
although the resolution is not sufficient to rule out steps on the
scale of \(\approx \SI{10}{\kelvin}\).

\subsection{Longitudinal-field measurements}
To elucidate the dynamic response, a series of LF measurement were
performed at both \(T = \SI{200}{\kelvin}\) and
\(T = \SI{350}{\kelvin}\) by applying a series of external magnetic
fields (up to $B = \SI{0.5}{\tesla}$) along the direction of the muon
spin. As the field magnitude increases the Zeeman term in the muon's
Hamiltonian dominates, and the muon spin is pinned along its initial
direction. Time-varying local fields can then cause a muon spin flip
and relax the asymmetry. This state of affairs allows us to
investigate the magnetic-moment dynamics in the two phases by fitting
the results to a series of exponential functions, quantifying the
relaxation due to the fluctuating magnetic fields. This is appropriate
even in the high \(T\) limit, since the applied field rapidly quenches
the Gaussian relaxation, leaving residual exponential relaxation
reflecting electronic dynamics. The model used is therefore
\begin{equation}
  \label{eq/lf_model}
  A(t)  = A_{\mathrm{R}}^{\mathrm{LF}}\exp(-\lambda^{\mathrm{LF}} t) + A_{\mathrm{B}}^{\mathrm{LF}},
\end{equation}
where we fix the parameter
\(A_{\mathrm{B}}^{\mathrm{LF}} = \SI{10}{\percent}\) to simplify the
fitting procedure. The value of the relaxation rate
\(\lambda^{\mathrm{LF}}\) is also shown in \cref{fig/mag_params}(c)
for both temperatures. We see that only the low-temperature
measurements show a decrease with increasing magnetic field. This
relationship can be fitted to the Redfield formula \cite{blundell2022}
\begin{equation}
  \label{eq/redfield_formula}
  \lambda^{\mathrm{LF}} = \frac{2\Delta^{2}\nu}{\nu^{2}+ \gamma_{\mu}^{2}B_{0}^{2}} + \lambda_0,
\end{equation}
where \(\Delta\) is the fluctuating amplitude
(\(\Delta^2 / \gamma_{\mu}^2 = {\ev*{(\delta B)^2}}\)), \(\nu\) is
fluctuation rate (related to \(\tau = \nu^{-1}\) the correlation time
between changes), \(B_0\) is the applied external field and
\(\lambda_0\) is an offset accounting for the component of the
relaxation not reduced by the external field. (Such an offset is often
observed in dynamically-fluctuating molecular systems
\cite{lancaster2004}). This gives values of
\(\nu = \SI{66 \pm 12}{\mega\hertz}\) and
\(\Delta = \SI{3.1 \pm 0.3}{\per\micro\second}\) for the parameters.
In the high-temperature phase a very small relaxation rate is observed
in applied field, confirming that the ZF relaxation is caused by
static nuclear moments, which are unable to cause the required spin
flips. On the other hand, the successful description of the
low-temperature relaxation parameters with the Redfield formula
confirms that the muon-spin relaxation is caused by randomized
electronic moments with dynamics in the fast-fluctuation limit.

\section{Muon site analysis}

First-principles calculations allow us to compute candidate muon sites and gain an insight into how the muon probes materials. This method (DFT+\(\mu\) \cite{lancaster2018,huddart2021,blundell2013}) involves performing a geometry optimisation calculation of the structure with an additional reduced-mass hydrogen atom representing a bare muon (\(\mathrm{Mu}^{+}\)) or a muonium (\(\mathrm{Mu}^{0}\)) atom. The initial implantation site is chosen randomly with the constraint that initial positions must be a minimum distance of \(\SI{1}{\angstrom}\) from the atoms and \(\SI{0.5}{\angstrom}\) from the other sites. The simulations were run on a \(\num{8.6} \times \num{9.9} \times \SI{21.4}{\angstrom}\) unit cell of the crystal structure of HbimDTDA using the CASTEP code \cite{clark2005} with files generated by the MuFinder program \cite{huddart2021}, producing 30 candidate sites for each phase. The relaxed unit cells were analysed first by considering the distortions to the atomic positions caused by the muon, which in this case are minimal between atoms of the same molecule but more considerable between molecules, with a maximum radial displacement of \(\SI{\sim 1.0}{\angstrom}\), especially for the sites \(\mathsf{\bar{H}}\)~/~\textcolor[HTML]{8C564B}{$\blacksquare$} and \(\mathsf{H}\)~/~\textcolor[HTML]{E377C2}{$\blacksquare$} described below. Muon sites corresponding to different relaxed structures were compared by using the vector between the site and closest atom to position the muons in a undistorted cell. Finally the symmetry of the crystal was used to move all the sites to the same molecule and nearby sites (\(d < \SI{1}{\angstrom}\)) were grouped together by averaging their positions. All the muonium site simulations were also repeated for a bare muon, by not adding an extra electron to the system for the muon. This gave similar results, so that sites were matched with the muonium ones by assuming that ones closer than \(\SI{0.5}{\angstrom}\) are equivalent. All sites were realised in both cases with the exception of a single muon site (denoted \(\mathsf{S_1}\)~/~\textcolor[HTML]{FF7F0E}{$\blacksquare$} below) which was not found in the high-temperature phase (details can be found in the Supplemental Material \cite{supplemental_meterial}).

\begin{figure}
\centering
\begin{tabular}{c@{\hspace{0.1cm}}c@{\hspace{0.4cm}}c@{\hspace{0.1cm}}c@{\hspace{0.4cm}}c@{\hspace{0.1cm}}c@{\hspace{0.4cm}}c@{\hspace{0.1cm}}c}
H & \textcolor[HTML]{CDCDCD}{$\blacksquare$} &
C & \textcolor[HTML]{838383}{$\blacksquare$} &
N & \textcolor[HTML]{617CCF}{$\blacksquare$} &
S & \textcolor[HTML]{D1D161}{$\blacksquare$}
\end{tabular}
\\[0.5cm]
\begin{subfigure}{1.0\linewidth}
\centering
\includegraphics[width=0.6\columnwidth]{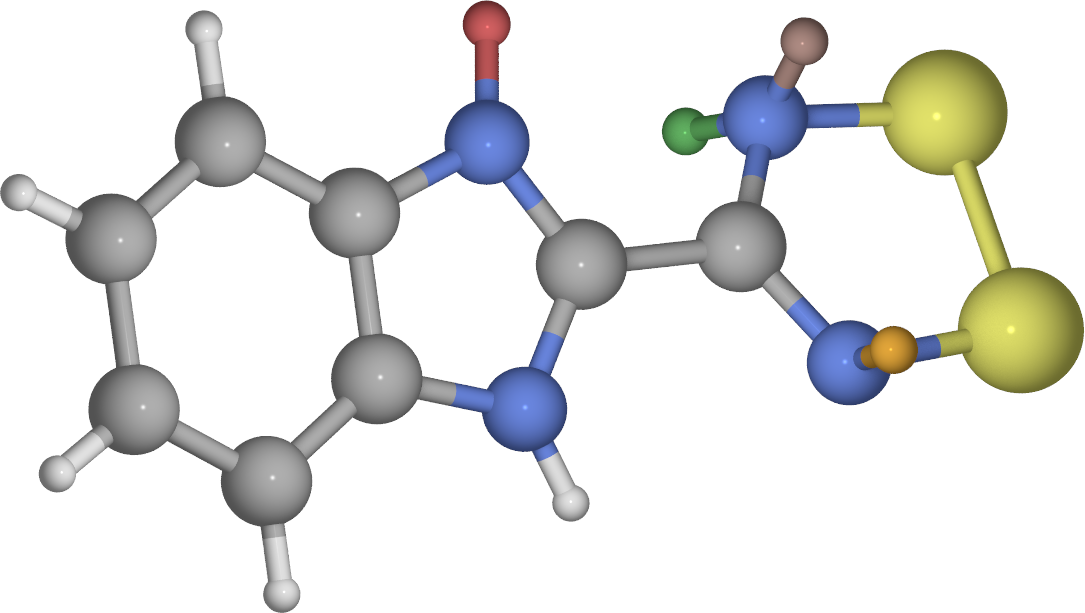}
\caption{\SI{100}{\kelvin} (top)}
\label{fig/sites_100k_top}
\end{subfigure}
\\[1ex]
\begin{subfigure}{1.0\linewidth}
\centering
\includegraphics[width=0.6\columnwidth]{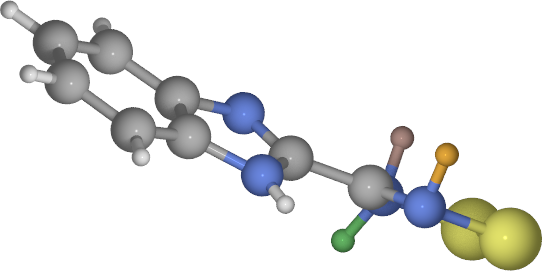}
\caption{\SI{100}{\kelvin} (side)}
\label{fig/sites_100k_side}
\end{subfigure}
\\[1ex]
\begin{subfigure}{1.0\linewidth}
\centering
\includegraphics[width=0.6\columnwidth]{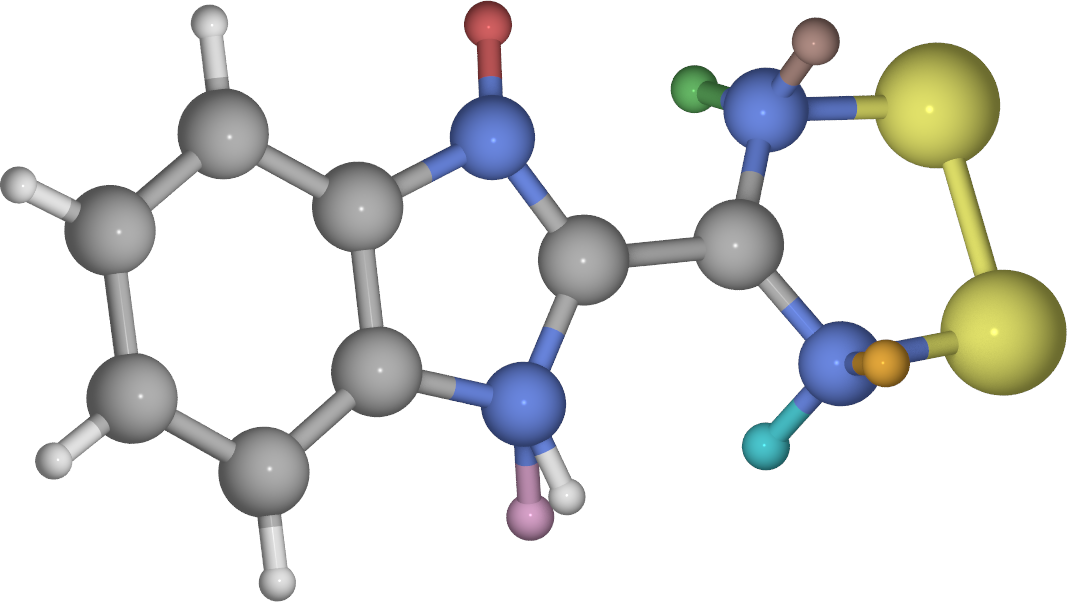}
\caption{\SI{340}{\kelvin} (top)}
\label{fig/sites_340k_top}
\end{subfigure}
\\[1ex]
\begin{subfigure}{1.0\linewidth}
\centering
\includegraphics[width=0.6\columnwidth]{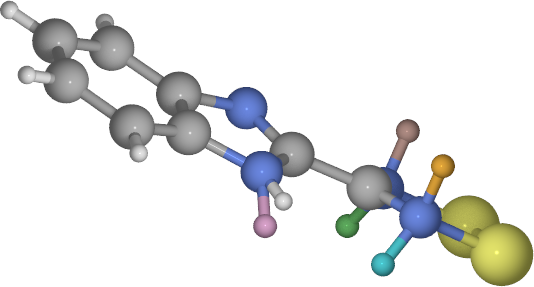}
\caption{\SI{340}{\kelvin}  (side)}
\label{fig/sites_340k_side}
\end{subfigure}
\caption{Diagrams showing the main sites for muonium at (a,b) \SI{100}{\kelvin}, with three low-energy sites (\textcolor[HTML]{FF7F0E}{$\blacksquare$}, \textcolor[HTML]{2CA02C}{$\blacksquare$} and \textcolor[HTML]{8C564B}{$\blacksquare$}) and a slightly higher-energy site (\textcolor[HTML]{D62728}{$\blacksquare$}) (c,d) \SI{340}{\kelvin}, with two new lower energy sites (\textcolor[HTML]{E377C2}{$\blacksquare$} and \textcolor[HTML]{17BECF}{$\blacksquare$})}
\label{fig/muon_sites}
\end{figure}

The positions of the calculated candidate muon sites are shown in \cref{fig/muon_sites} (only for the case of \(\mathrm{Mu}^{0}\) but the others are similar) and their respective energies are listed in \cref{tbl/site_energies}. Energies are given relative to the lowest-energy site for each column. The similarity in energy between the candidate sites in each class suggests that we might expect each of them to be realised. We first find a set of candidate sites common to both structures close to the nitrogen atoms in the sulphur-containing rings. Two of them (\(\mathsf{S_1}\)~/~\textcolor[HTML]{FF7F0E}{$\blacksquare$} and \(\mathsf{S_2}\)~/~\textcolor[HTML]{2CA02C}{$\blacksquare$}) are located \emph{outside} the region between rings in adjacent chains (see shaded area in \cref{fig/molecule:chain_low}), with the second being \emph{closer} to the atoms which form the contact bond between chains in the high temperature phase (see \cref{fig/molecule:chain_high}). Another site (\(\mathsf{S_3}\)~/~\textcolor[HTML]{D62728}{$\blacksquare$}) is located \emph{inside} the region but \emph{away} from the contact sulphur atoms, which might explain why it has a similar energy at the lower temperature but is higher in energy at \(\SI{340}{\kelvin}\). The other low-temperature site (\(\mathsf{\bar{H}}\)~/~\textcolor[HTML]{8C564B}{$\blacksquare$}) has the muon attached to the non-hydrogenated nitrogen atom in the central ring and is higher in energy for muonium but the lowest energy site in the case of the bare muon.

Apart from these common sites, for the \(\SI{340}{\kelvin}\) structure we also find two new lower-energy sites. One (\(\mathsf{H}\)~/~\textcolor[HTML]{E377C2}{$\blacksquare$}) is found sharing the nitrogen atom with a hydrogen atom in the central ring (see \cref{fig/molecule:chemfig}) and the other (\(\mathsf{S_4}\)~/~\textcolor[HTML]{17BECF}{$\blacksquare$}) is again attached to one of the nitrogen atoms in the sulphur-containing ring, but in this case is \emph{inside} the inter-ring region and \emph{closer} to the contact atoms. To explain the presence of the new sites we note that the main difference between the two structural phases is the presence of the pancake bonds between the sulphur rings in the lower-temperature state and the relative position of the chain. The breaking of these bonds at higher temperature seems to make the new positions available.

\begin{table}[h]
\centering
\begin{tabular}{{l@{\ }lrrrr}}
\toprule
\multicolumn{2}{c}{\multirow{3}{*}{Site}} & \multicolumn{4}{c}{Energy (eV)} \\
\cmidrule(l){3-6}
\multicolumn{2}{c}{} & \multicolumn{2}{c}{\(\textsf{Mu}^{0}\)} & \multicolumn{2}{c}{\(\textsf{Mu}^{+}\)} \\
\cmidrule(l){3-4} \cmidrule(l){5-6}
\multicolumn{2}{c}{} & \multicolumn{1}{c}{100\,K} & \multicolumn{1}{c}{340\,K} & \multicolumn{1}{c}{100\,K} & \multicolumn{1}{c}{340\,K} \\
\midrule
\(\mathsf{S_1}\) & (\textcolor[HTML]{FF7F0E}{$\blacksquare$}) & 0.01 & 0.06 & 0.43 & - \\
\(\mathsf{S_2}\) & (\textcolor[HTML]{2CA02C}{$\blacksquare$}) & 0.01 & 0.17 & 0.38 & 0.93 \\
\(\mathsf{S_3}\) & (\textcolor[HTML]{D62728}{$\blacksquare$}) & 0.12 & 0.14 & \textbf{0.00} & \textbf{0.00} \\
\(\mathsf{\bar{H}}\) & (\textcolor[HTML]{8C564B}{$\blacksquare$}) & \textbf{0.00} & 0.06 & 0.37 & 0.24 \\
\(\mathsf{H}\) & (\textcolor[HTML]{E377C2}{$\blacksquare$}) & - & 0.05 & - & \textbf{0.00} \\
\(\mathsf{S_4}\) & (\textcolor[HTML]{17BECF}{$\blacksquare$}) & - & \textbf{0.00} & - & 0.23 \\
\bottomrule
\end{tabular}
\caption{Table comparing the energies of the unit cell with the muon at the different bare muon (\(\textsf{Mu}^{+}\)) and muonium (\(\textsf{Mu}^{0}\)) sites calculated using DFT, and given relative to the lowest energy found in each column.}
\label{tbl/site_energies}
\end{table}

\section{Discussion}

Conventionally we assume that a bare (or diamagnetic) muon spin
couples to the local magnetic field in a material, and probes the
local field distribution without causing an appreciable perturbation.
The relevant muon sites from the previous section would then be the
bare ones. The low-temperature regime of this material, which is
thought to be formed from singlet spins, was previously suggested to
be diamagnetic on the basis of bulk susceptibility measurements.
However, if the muon takes the form of an unperturbing, diamagnetic
probe, then the low-temperature relaxation cannot simply be explained
by the presence of highly-dilute magnetic impurities in a diamagnetic
background, since the Redfield behaviour observed relies on the
presence of a dense array of magnetic moments that rapidly fluctuate
in time. It might therefore be unlikely that the material can be
characterised as being non-magnetic in this regime, but rather there
are fluctuating moments of sufficient density to be approximated as
giving rise to a Gaussian distribution of fields at any instant. We
distinguish the local magnetic field distribution in the low
temperature phase, featuring this distribution of moments fluctuating
in the fast fluctuation limit, from that in the high-temperature
regime, which likely comprises a denser distribution of moments, with
a far greater characteristic fluctuation rate.

Since the muon is a local probe, the transition we observe likely
reflects muons locally detecting the switching of nearby clusters of
molecules in the sample. A cluster in the low-temperature state giving
an exponential relaxation and one in the high-temperature state a
Gaussian one. The stretched exponential used in the intermediate
regime then models the sum of contributions, whose relative size
varies with temperature. We note that our results in this system
resemble those measured in spin-crossover systems based on iron (II)
ions which show a crossover between a low-spin (\(S = 0\)) state at
low temperature and a high-spin (\(S = 2\)) state at high temperature
\cite{blundell2003,blundell2004}. In those materials the muon spectra
were also fitted to a stretched-exponential function with
\(\beta < 1\) in the low temperature, low-spin configuration and
\(\beta\) approaching \(\beta = 2\) at high temperature. It was
suggested that the relaxation at low temperature reflected an
incomplete crossover, with some spins remaining in the high-spin
configuration at low temperature and forming a very dilute
distribution leading to root-exponential relaxation \cite{uemura85}. A
similar picture could be the case here, with any regions that avoid
the low-temperature structural transition giving rise to a
distribution of disordered spins, causing the observed relaxation.
However, the fact that we find \(\beta \lesssim 1\) suggests that the
density of moments in our system at low temperature is greater than
the highly-dilute one that would be expected to give rise to
\(\beta \approx 0.5\), which was the value observed in some of the
low-temperature phases of the iron-based spin-crossover systems
\cite{blundell2004}.

Another possibility which could account for our data is that the
muon's sensitivity to the magnetism in the low-temperature, singlet
state is caused by a perturbation the muon makes to the system, as was
suggested to be the case in molecular spin-ladder materials
\cite{lancaster2018}. This might involve the bare, charged muon
causing a local distortion to the nearest spin singlet, or that the
sensitive species is derived from muonium, whose extra electron is
involved in causing the necessary distortion. The muon, along with its
local distortion, would then become the sensitive species, whose
interactions give rise to the observed relaxation. We note that the
observed fluctuation amplitude \(\Delta\) in this regime corresponds
to the magnetic field from an electron spin around $\approx 6$~\AA\
from a muon, providing a rough length scale for the interaction. If
this is the case, then the material could adopt a fairly uniform
singlet ground state with few additional intrinsic magnetic
impurities. However, even in this case, the transition to a regime of
large, dense magnetic moments at high temperature would continue to
allow the muon to faithfully probe the magnetic switching transition.

Finally, the difference in the low-energy muon sites in this
material's two structural phases is a noteworthy feature, as the
difference in sites in different states of a system has not been
discussed previously in materials of this type. Since we find a range
of muon sites in this system with very similar energy, we would expect
the muons to sample a range of internal magnetic fields. Although both
a bare muon and muonium allow several different low-energy candidate
sites in the two temperature regimes, owing to the range of fields
probed, the two cases are unlikely to lead to cause significant
differences in to the measured spectra. However, the observation of
new sites becoming available after a structural change likely applies
well beyond this material.

Important questions remain about the nature of the phase transition in
this system, particularly related to the broadness of the transition
compared to the width of the hysteretic region. Inspection of the
magnetic susceptibility data suggests the presence of steps in the
response. Indeed, tracking the structural component of the transition
as a function of \(T\) by powder x-ray diffraction also suggests a
stepwise progression, with reflections consistent with the high
temperature phase appearing over a range of temperatures. There has
been recent interest in the possibility of realizing the devil's
staircase structure in such systems \cite{trzop2016}, where step-like
transitions between the spin states have been observed.

\section{Conclusion}

Muon-spin relaxation measurements, paired with muon-site analysis,
have allowed us to probe the hysteretic magnetic switching behaviour
of HbimDTDA from a local perspective. We identify a hysteresis width
of \(\Delta T \approx \hystWidth\), centred on \(T = \hystCenter\).
The low-temperature state gives rise to muon-spin relaxation which is
well described by a model that assumes a dense arrangement of
disordered, dynamically-fluctuating moments. The structural transition
causes the muon sites in the two regimes to differ. However, in a
chemically-complex material such as this, a large number of sites of
similar energy occur in both regimes, with the result that we expect
the muon to faithfully probe the system across the transition.
Although this latter feature of differing muon sites in different
structural regimes has yet to be widely investigated, it is possible
that it is a general feature that we should expect in numerous
systems.

\section{Acknowledgments}

Part of this work was performed at the STFC-ISIS facility and we are
grateful for provision of beamtime and access to the SCARF Computer
cluster. We are also grateful for computational support provided by
both Durham Hamilton HPC and by the UK national high performance
computing service, ARCHER2, for which access was obtained via the UKCP
consortium and funded by EPSRC.
AHM is grateful to STFC and EPSRC for the provision of a studentship.
For the purpose of open access, the authors has applied a Creative
Commons Attribution (CC BY) licence to any Author Accepted Manuscript
version arising. Research data from this project will be made
available via Durham Collections.

\printcredits

\bibliography{refs}



\end{document}